# A particle–water based model for water retention hysteresis

Y. GAN\*, F. MAGGI\*, G. BUSCARNERA† and I. EINAV\*

A particle–water discrete element based approach to describe water movement in partially saturated granular media is presented and tested. Water potential is governed by both capillary bridges, dominant at low saturations, and the pressure of entrapped air, dominant at high saturations. The approach captures the hysteresis of water retention during wetting and drainage by introducing the local evolution of liquid–solid contact angles at the level of pores and grains. Extensive comparisons against experimental data are presented. While these are made without the involvement of any fitting parameters, the method demonstrates relative high success by achieving a correlation coefficient of at least 82%, and mostly above 90%. For the tested materials with relatively mono-disperse grain size, the hysteresis of water retention during cycles of wetting and drainage has been shown to arise from the dynamics of solid–liquid contact angles as a function of local liquid volume changes.

## INTRODUCTION

Water retention in partially saturated soil is fundamental in soil mechanics (Loret & Khalili, 2002; Nuth & Laloui, 2008; Sheng *et al.*, 2008; Buscarnera & Einav, 2012) and has important implications in soil ecohydrology (Laio *et al.*, 2001) as well as nutrient cycle and microbial ecology (Crawford *et al.*, 2005; Maggi *et al.*, 2008). Of particular interest in recent decades is water management in agriculture (Blonquist *et al.*, 2006; Heinse *et al.*, 2007; Gu *et al.*, 2009), contaminant flow (Dury *et al.*, 1998; Gerhard & Kueper, 2003; Culligan *et al.*, 2004; Serrano *et al.*, 2011), bio- and phytoremediation (Salt *et al.*, 1998; Pilon-Smits, 2005) and water treatment (Magnuson & Speth, 2005; Schideman *et al.*, 2006).

The water retention relation between water potential and degree of saturation is typically used to define the hydraulic behaviour of soils. A key element of this relation is the hysteresis observed during cycles of wetting and drainage. Water retention hysteresis in soils may be ascribed to geometric or ink-bottle effects, differences in contact angles during wetting and drying, entrapment of a non-wetting phase (gas, oil, etc.) and the shrinking and swelling of pores (Morrow, 1976; Tindall *et al.*, 1999; Lappalainen *et al.*, 2008; Pereira & Arson, 2013).

The overall behaviour of partially saturated soils depends on the solid skeleton, the way liquids and gases are connected, and the way forces interact along interfaces. Soil properties can be characterised by the pore size distribution (Revil & Cathles, 1999; Gili & Alonso, 2002) and by the distributions of grain sizes and shapes (Blonquist *et al.*, 2006; Heinse *et al.*, 2007). The interfaces between the solid, liquid and gaseous phases can be characterised by the grain surface profiles, their hydrophilicity and thermodynamic properties (Blake & Haynes, 1969; Bachmann *et al.*, 2003; Goebel & Bachmann, 2004; Lourenço *et al.*, 2012; Russell & Buzzi, 2012).

Existing phenomenological models parameterise the water retention curve using parameters that do not necessarily have a direct physical meaning (Richards, 1931; van Genuchten, 1980; Heinse *et al.*, 2007). For example, two sets of parameters are to be determined for the van Genuchten equation to reproduce the water retention curve during drainage and wetting, yet there is no consensus on what these two sets of parameters mean. What may be relevant is to establish an explanation that would bypass such phenomenology and thus highlight what actually happens at the grain scale and at the interfaces between solids, liquids and gases.

Numerical modelling of water flow and the resulting retention is feasible with the aid of discrete element methods (DEMs) (Cundall & Strack, 1979), but this has not yet revealed water retention hysteresis, at least not in a quantitatively successful way. Nevertheless, DEMs have been used to describe inter-granular capillary interactions at low water saturations (i.e. for continuous gas and discontinuous liquid, or the pendular state) (Gili & Alonso, 2002; Liu & Sun, 2002; Jiang *et al.*, 2004; Soulié *et al.*, 2006a, 2006b; Scholtès *et al.*, 2009a, 2009b; Gras *et al.*, 2011; Schwarze *et al.*, 2013). An extension of the DEM was proposed to incorporate intermediate saturations (i.e. for continuous gas and liquid phases, or the funicular state) (Zeghal & El Shamy, 2004; Di Renzo & Di Maio, 2007; Chareyre *et al.*, 2012) and full saturation (i.e. for continuous liquid and discontinuous gas, or the capillary state). However, only the coupling between these approaches would allow one to capture in full the retention curve of partially saturated soils during wetting and drying. Moreover, the coupling of these requires a scheme that extends beyond the capabilities of each of these individual frameworks and that allows for model testing and validation against experimental observations of water retention, including hysteresis features.

The aim of this study was to develop a new DEM model for partially saturated granular materials, which is coupled with a newly developed homogenisation scheme designed to connect local and averaged water potentials. The DEM model is based on the description of local water movement between neighbouring cells, each containing a single grain. The homogenisation scheme is used to determine the water retention curve, hence the relation between the averaged water potential against the averaged degree of saturation. The proposed homogenisation scheme is capable of describing partial saturation and is specifically used to detect hydraulic hysteresis during wetting and drainage. The proposed DEM model was tested for hysteresis cycles against three sets of experiments incorporating variable grain size distribution, surface tension between phases and contact angles.

## GRAIN-SCALE PROPERTIES

Grain-scale properties are defined in relation to tessellated cells. The boundaries of the $i$th tessellated cell are obtained using a modified Voronoi tessellation method (Rycroft *et al.*, 2006), where the $i$th cell volume $V_i^{cell}$ is centred around the corresponding grain (Fig. 1). The total volume $V^{total}$ is equal to the sum of $V_i^{cell}$, and each cell $i$ is associated with a cell solid volume $V_i^{grain}$, water volume $V_i^{water}$ and void volume $V_i^{void} = V_i^{cell} - V_i^{grain}$. Hence, the volumes at the scale of the domain are

$$V^{void} = \sum_i V_i^{void}$$

$$V^{water} = \sum_i V_i^{water}$$

$$S_r = \frac{V^{water}}{V^{void}}$$

The degree of saturation, porosity and water content of the $i$th cell are defined by

$$S_{r,i} = \frac{V_i^{water}}{V_i^{void}} \quad (1a)$$

$$\phi_i = \frac{V_i^{void}}{V_i^{cell}} \quad (1b)$$

$$\Theta_i = \frac{V_i^{water}}{V_i^{cell}} \quad (1c)$$

Capillary pressure due to the presence of a water meniscus between two grains adds to inter-granular forces in this DEM, which is calculated using the solution to the Young–Laplace equation as

$$F_{ij}^{cap} = F^{cap}(R_i, R_j, \delta_{ij}, V_{ij}^{br}, \gamma, \theta) \quad (2)$$

where $R_i$ and $R_j$ are the radii of grains $i$ and $j$, $\delta_{ij}$ is the gap distance, $V_{ij}^{br}$ is water volume of the capillary bridge, $\gamma$ is the surface tension and $\theta$ is the contact angle between water and grain. In this study, the approximate solution provided by Soulié *et al.* (2006a) is used.

An important new aspect of the current method acknowledges the evolution of the contact angle $\theta$ between physically meaningful minimum and maximum values ($\theta_{min}$ and $\theta_{max}$) when the water bridge recedes or expands, respectively. The rate of change in contact angle is assumed to take the form

$$\dot{\theta} = \alpha \frac{\dot{V}_{ij}^{br}}{R_{eff}^3} \quad \theta_{min} \leq \theta \leq \theta_{max} \quad (3)$$

where $\dot{V}^{br}$ is the rate of change of water volume of this capillary bridge, $\alpha$ is a non-dimensional coefficient controlling the dependence of the contact angle on the change of local water volume and $R_{eff}$ is the effective radius, given by $R_{eff} = 2R_iR_j/(R_i+R_j)$. For a given pair of liquid and solid surfaces, the actual values of $\theta_{min}$ and $\theta_{max}$ depend not only on the chemical properties of the solid grain, but also on its surface morphology. The dynamics of $\theta$ between $\theta_{min}$ and $\theta_{max}$ is later shown to help us capture the hysteresis response during wetting and drainage (see Fig. 2).

Above a local critical degree of saturation $S_r^c$, pressure due to entrapment of gas bubbles is introduced. The local critical degree of saturation $S_r^c$ may depend on the grain's morphology through its fractal surface and it may be related to the air entry value in terms of saturation (Cho & Santamarina, 2001), but details will not be discussed here.

Assuming that the moles of gas $n$ and temperature $T$ are constant, according to the ideal gas law, $u_p^a V_p^a = nRT$ (the superscript $a$ denotes the air–gas phase and subscript $p$ denotes an association to a particle-cell), and the relative air pressure can be calculated as

$$\Delta u_p^a = u_p^a - u^a = \begin{cases} 0 & \text{if } S_r \leq S_r^c \\ \frac{S_r - S_r^c}{1 - S_r} u^a & \text{if } S_r > S_r^c \end{cases} \quad (4)$$

where $u^a$ and $u_p^a$ are the atmospheric and particle-cell air pressures respectively. The entrapped air volume is defined through $S_r^c$, as $V_0^{air} = (1 - S_r^c)V^{void}$. Note that, according to equation (4), $S_r$ avoids approaching unity with increasing relative air pressure due to the denominator approaching zero, with liquid tending to move to neighbouring cells.

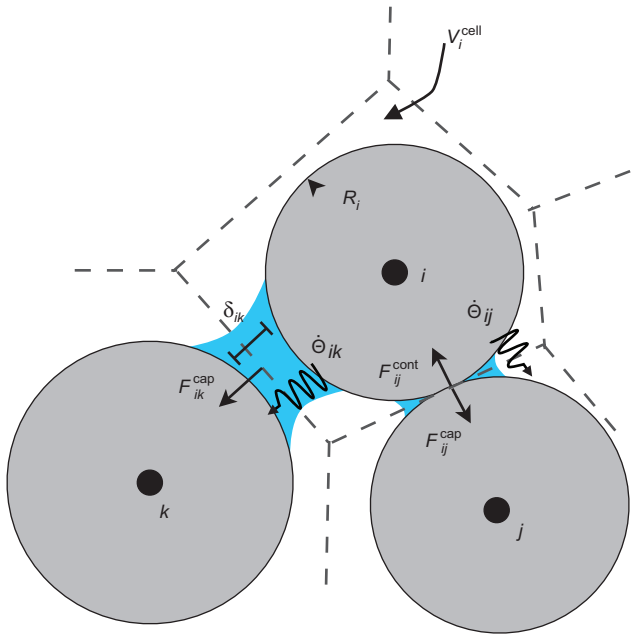

**Fig. 1.** Schematic illustration of three grains (grain centres indicated by $i$, $j$ and $k$) with connecting capillary bridges. The space was divided using Voronoi tessellation, and water volumes within the cells vary by mass flux through cell boundaries as $\dot{\Theta}_{ij}$ and $\dot{\Theta}_{ik}$

## HOMOGENISATION SCHEME

A homogenisation scheme is proposed to determine collective water potential in terms of local quantities. According to a common definition in partially saturated soil mechanics, the effective stress $\sigma'_{ij}$ is related to the net

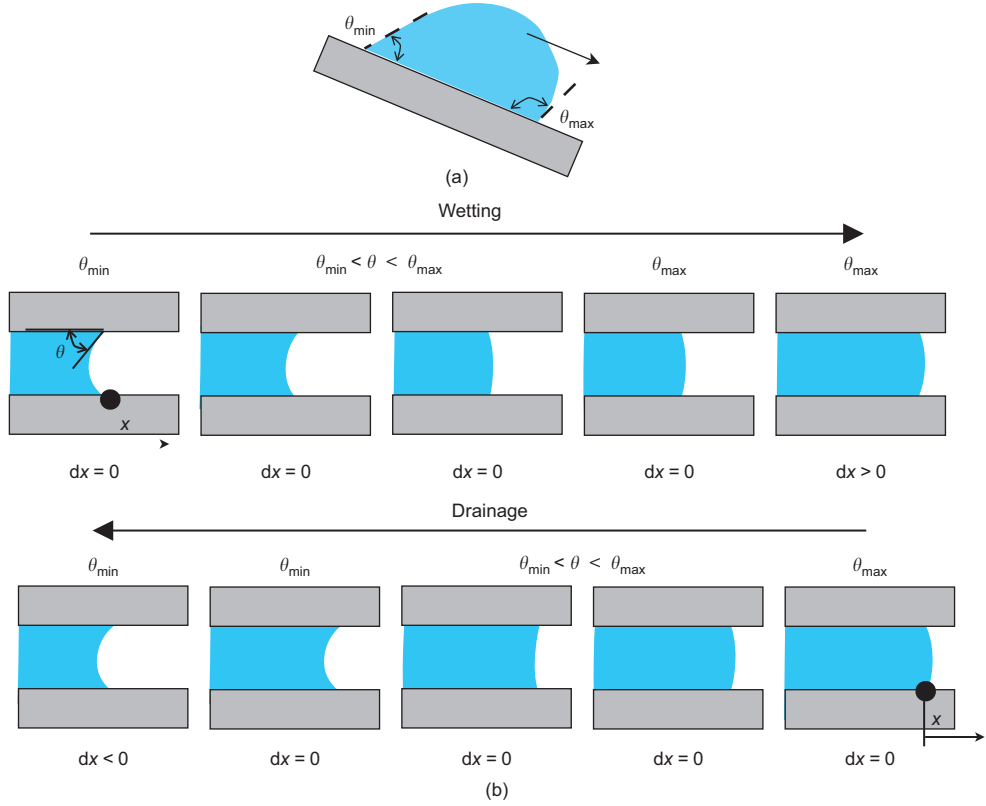

**Fig. 2.** Schematic illustration of contact angle dynamics. (a) Contact angle limits, shown for a water drop sliding along a tilted surface. (b) Contact angle dynamics during wetting and drainage, with d$x$ indicating the displacement of the three-phase contact line

stress tensor $\sigma_{ij}^{net} = \sigma_{ij} - u^a \delta_{ij}$, water pressure $u^w$ and atmospheric air pressure $u^a$ using (Houlsby, 1997)

$$\sigma_{ij}^{net} = \sigma'_{ij} - S_r \psi \delta_{ij} \tag{5}$$

where $\psi = u^a - u^w$ is the water potential and $\delta_{ij}$ is the Kronecker delta. A more general definition of the effective stress is to use the weighting parameter $\chi(S_r)$ for the water potential (Gray & Schrefler, 2001; Coussy et al., 2010; Nikooee et al., 2012). Note that, for saturations above $S_r^c$, the particle-cell air pressure $u_p^a$ is not the atmospheric air pressure $u^a$ at saturation (see equation (4)).

Scholtès et al. (2009b) separate the net stress tensor into effective and capillary stress tensors

$$\langle\sigma\rangle_{ij}^{net} = \langle\sigma\rangle'_{ij} + \langle\sigma\rangle_{ij}^{cap} = \frac{1}{V}\left(\sum_c F_i^{cont} x_j + \sum_c F_i^{cap} x_j\right)$$

where $x$ is the centre-to-centre vector between two neighbouring grains, with the contact and capillary forces defined by $F^{cont}$ and $F^{cap}$, respectively. The expression $\langle\cdot\rangle$ indicates a volume average. A more general equilibrium equation should introduce forces due to relative air pressure

$$\begin{aligned}\langle\sigma\rangle_{ij}^{net} &= \langle\sigma\rangle'_{ij} + \langle\sigma\rangle_{ij}^{cap} + \langle\sigma\rangle_{ij}^{air}\\ &= \frac{1}{V}\left(\sum_c F_i^{cont} x_j + \sum_c F_i^{cap} x_j + \sum_p \Delta u_p^a V_p^{cell} \delta_{ij}\right)\end{aligned} \tag{6}$$

where we correlate the effective stress tensor $\langle\sigma\rangle'_{ij}$ to the inter-granular contact forces, the capillary stress tensor $\langle\sigma\rangle_{ij}^{cap}$ to the capillary forces and the intrinsic particle-cell air pressure $\langle\sigma\rangle_{ij}^{air}$ to the relative air pressures. A comparison of equation (5) with equation (6) shows the following trace of stress tensors

$$\langle\sigma\rangle_{ii}^{cap} + \langle\sigma\rangle_{ii}^{air} = -3S_r \psi$$

Thus, the water potential $\psi$ can be stated in terms of homogenised quantities

$$\psi = -\frac{1}{3S_r V}\left(\sum_c F_i^{cap} x_i + \sum_p \Delta u_p^a V_p^{cell}\right) \tag{7}$$

The second term of equation (7) is negligible at low saturation, while the first is negligible near saturation. Note that this homogenisation scheme may be included in systems with more than one component per phase (Maggi, 2012).

The rate of change of the water content (in the cell scale $\Theta_i = V_i^{water}/V_i^{cell}$) can be calculated with the Richards equation (Richards, 1931)

$$\frac{\partial \Theta}{\partial t} = \nabla[K(\Theta)\nabla(\psi + \rho g z)] \tag{8}$$

where $K(\Theta)$ is the hydraulic conductivity and $z$ is the vertical position. The (cell) volume integral of equation (8) can be written as the circuitation on the cell boundary using the Green's theorem as

$$\int \frac{\partial \Theta}{\partial t} dV = -\oint \boldsymbol{\Phi}\cdot\boldsymbol{n}\, dS \tag{9}$$

where $\boldsymbol{n}$ is the surface normal, $\boldsymbol{\Phi} = -K(\Theta)\nabla(\psi + \rho g z)$ is the water flux and $\psi$ is determined as in equation (7) for each cell. The change of local water volume can be calculated using the individual fluxes between its neighbouring grains, through the discrete version of equation (9)

**Table 1.** Model parameters used in the simulations

| Variable | Value |
| --- | --- |
| Grain size, $d$ | $1\cdot 0 \pm 0\cdot 05$ |
| Porosity, $\phi$ | 37% |
| Surface tension, $\gamma$ | 0·01 |
| Hydraulic conductivity, $K_0$ | 5·0 |
| Contact angle coefficient, $\alpha$ | 10·0 |
| Saturation for air entrapment, $S_r^c$ | 0·98 |
| Atmospheric air pressure, $u^a$ | 0·01 |
| Elastic modulus of grains, $E/(1-v^2)$ | 1000 |

$$\Delta V_i^{\text{water}} = V_i^{\text{cell}} \cdot \Delta \Theta = -\sum_j \Phi_{ij}^{\text{water}} A_{ij} \Delta t \qquad (10\text{a})$$

$$\Phi_{ij}^{\text{water}} = -\frac{K_0}{|\mathbf{x}_j - \mathbf{x}_i|}[\psi_j - \psi_i + \rho g(z_j - z_i)] \qquad (10\text{b})$$

It is assumed that changes of cell volume are negligible during each time iteration, i.e. $\Delta V_i^{\text{cell}} \approx 0$. Note also that it assumed that the hydraulic conductivity between adjacent cells is a function of the product between a constant conductivity per unit surface area $K_0$ and the conduction area $A_{ij}$ (the 'wet' area along a Voronoi interface), which corresponds to the cross-sectional area of the water bridge along the boundary between two cells. Compared with the hydraulic conductivity in equation (8), this approach allows us to set down a flow equation relatively simply while retaining the proportionality between ease of water movement and level of wetness (i.e. Darcian water velocity is greater as the medium is wetter). Within each cell, the total water volume is distributed to multiple existing water bridges (Soulié et al., 2006a); the time for redistribution within the cell is considered to be negligible compared with the time needed for water to move to connected cells. The capillary bridge will appear when two wetted grains are in contact and will rupture when the length of the bridge exceeds the rupture distance of $(1+\theta/2)(V_{ij}^{\text{br}})^{1/3}$ (Soulié et al., 2006a). The coalescence of water bridges at high saturations is not considered in this study.

## SIMULATION RESULTS AND DISCUSSION
### Initial and boundary conditions

A three-dimensional simulation domain consisting of 2500 grains is used. The grains are nearly mono-sized with 5% variation in diameter. The mechanical boundary conditions are periodic for grain dynamics. An initial porosity $\phi = 37\%$ is achieved by a stage of pre-compaction from a looser state. This corresponds to a close packing state for nearly mono-sized spherical grains. For this configuration, grain motion during the processes is negligible and the pore size distribution across the sample is nearly uniform. During simulations the domain volume is fixed (i.e. constant porosity). The hydraulic boundary conditions involve water inflow through a constant positive water pressure difference from the top during wetting and a water outflow through a constant negative water pressure difference from the bottom during drainage. Starting from an arbitrary degree of saturation (here, $S_r = 0\cdot 5$ for all grains), the wetting and drainage processes are applied to complete several scanning cycles between predefined maximum and minimum saturations for the whole simulation domain, until the sample provides a stable retention hysteresis. The collective water potential $\psi$ is calculated using equation (7). The parameters used in this study can be found in Table 1. The units in the simulations are not listed, but their magnitude is inspired by the later results related to the experimental data, where units are relevant. The contact force $\boldsymbol{F}^{\text{cont}}$ is calculated based on a Hertzian contact model with a friction coefficient of 0·5 (Gan & Kamlah, 2010).

### Hysteresis with contact angle dynamics

Three sets of numerical analyses were performed to show retention hysteresis predicted by the model and its dependency on the contact angle dynamics using

- various fixed contact angles
- the scanning window across various degrees of saturation
- various combinations of receding and advancing angles, $\theta_{\min}$ and $\theta_{\max}$.

Fixed contact angles, $\theta_{\min} = \theta_{\max} = 5°, 45°, 60°, 75°$ and $90°$, are first used to show the sensitivity of the water retention to $\theta$ in Fig. 3(a). The water potential is normalised as $\overline{\psi} = \psi \overline{d}/\gamma$, where $\overline{d}$ is the effective diameter for the system. This normalisation captures the effects of grain size and surface tension $\gamma$. In the simulations, the effective diameter is the mean diameter of the grains. By using a fixed contact angle $\theta$ for these mono-sized systems, the retention curves do not demonstrate hysteresis for the scanning between $S_r = 0\cdot 1$ and 0·9.

The next example involves contact angle dynamics between $\theta_{\min} = 5°$ and $\theta_{\max} = 60°$. There are three cases of hysteresis, for scanning between $0\cdot 1 \leq S_r \leq 0\cdot 9, 0\cdot 2 \leq S_r \leq 0\cdot 8$ and $0\cdot 4 \leq S_r \leq 0\cdot 6$ (Fig. 3(b)). The shape of the envelope of the scanning curves is determined by the dynamics of the contact angles. In the transition between drainage and wetting, the distribution of local contact angles determines the water retention relations between the water potential and the saturation (shown later in Fig. 8). The dynamics of the contact angle embedded in equation (3) holds the key for the model's ability to separate wetting and drainage branches of the hysteresis cycle. The shape of the hysteresis during transitions from drainage to wetting (and vice versa) is influenced by this equation, as observed by sensitivity to the parameter $\alpha$. The homogenised water potential $\overline{\psi}$ increases during drainage, due to an increase in $F^{\text{cap}}$ and a decrease in $S_r$, as reflected by equation (7).

Figure 3(c) shows the overall responses of three systems using different pairs of receding and advancing angles. The combination of $\theta_{\min}$ and $\theta_{\max}$ controls the shape and position of the retention hysteresis.

### Comparison with experiments

The model predictions were compared with experimental data from the existing literature to investigate three aspects that influence the overall retention hysteresis

- grain size (Heinse et al., 2007)
- surface tension (Dury et al., 1998)
- receding and advancing angles (Culligan et al., 2004).

Table 2 summarises the grain properties of these experiments, which were also directly used in the simulations.

In the first set of experiments, Heinse et al. (2007) measured the water retention hysteresis in granular materials composed of different sized aggregates. Figure 4 shows the comparison between the simulated hysteresis with the experimental data. For model-to-experiment comparison, the absolute matric potential $\psi/\rho g$ and water content $\Theta$ were used for samples having different grain size distributions (Figs 4(a)–4(c)). In Fig. 4(d), the experimental data are represented using a normalised potential $\overline{\psi} = \psi \overline{d}/\gamma$ and effective degree of saturation $S_r^{\text{eff}}(=(\Theta - \Theta_{\min})/(\Theta_{\max} - \Theta_{\min}))$. Use of the normalised $\overline{\psi}$ makes the evaluation independent of grain size. The correlation coefficients ($r_c$)

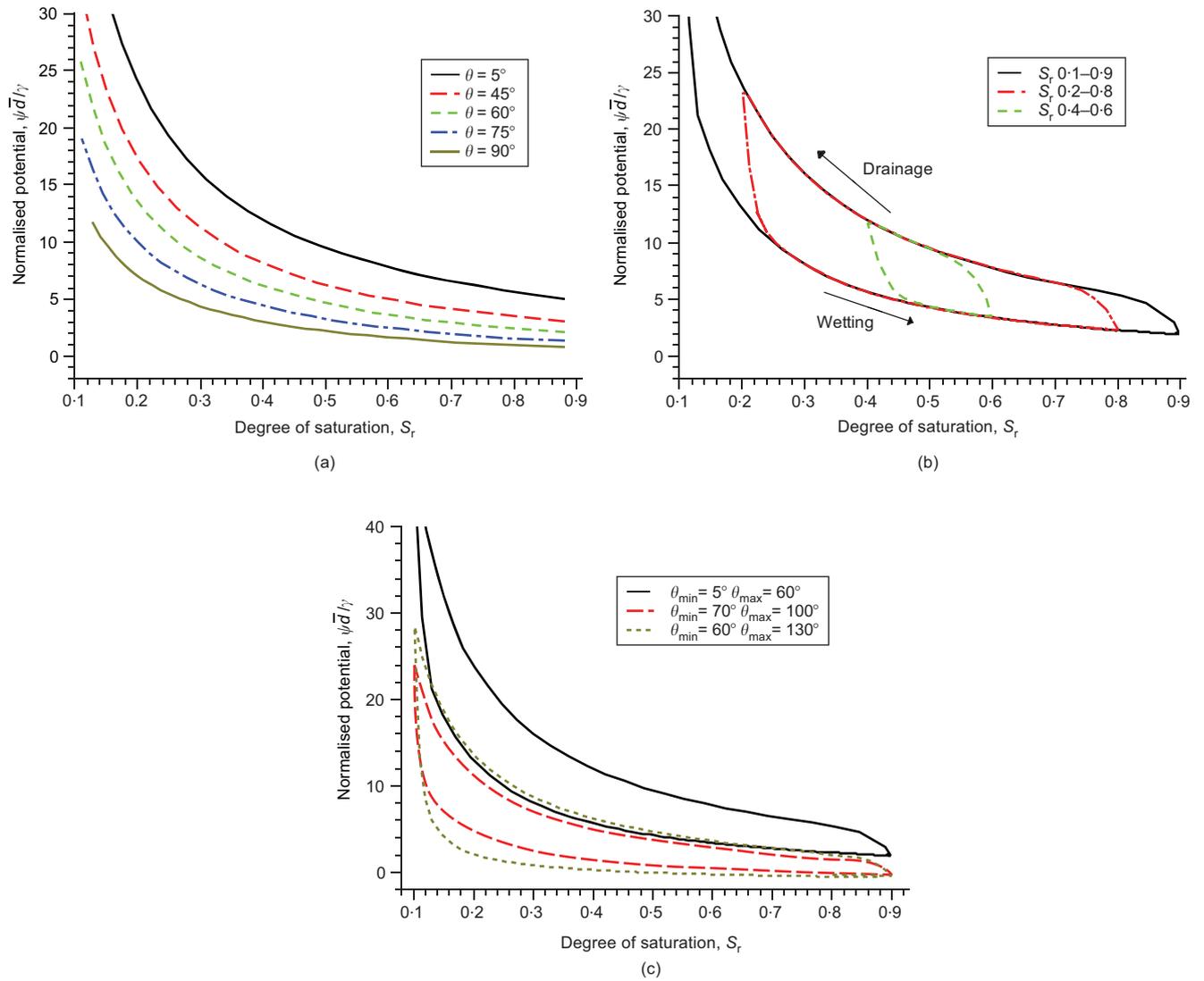

**Fig. 3.** Water retention curves: (a) fixed contact angles; (b) hysteresis using $\theta_{min} = 5°$ and $\theta_{max} = 60°$ for different scanning windows; (c) hysteresis for different pairs of receding and advancing angles. The arrows in (b) show the directions of drainage and wetting processes

between the model predictions and experimental data for wetting and drainage branches were calculated, separately, for all three cases. The correlation coefficients for the wetting branches range from 0·90 to 0·94, higher than those for the drainage branches, which range from 0·86 to 0·89. Note that this comparison did not involve any fitting parameters in the DEM simulations: this is therefore an *ab initio* comparison between numerical predictions and experiments. In general, the predictions for the wetting branch are better than those for the drainage branches.

**Table 2.** Properties of the experimental systems (properties listed were reported by the authors carrying out the tests, unless otherwise stated)

| Author | System | Contact angle $\theta$: degrees | Density $\rho$: kg/m³ | Surface tension $\gamma$: N/m | Diameter $d$: mm | Effective diameter $\bar{d}$: mm[a] |
|---|---|---|---|---|---|---|
| Heinse *et al.* (2007) | Profile | 5–60[b] | 1000 | 0·073 | 0·25–1·00 | 0·7 |
| | Turface | 5–60[b] | 1000 | 0·073 | 1·00–2·00 | 1·5 |
| | Mix | 5–60[b] | 1000 | 0·073 | 0·25–2·00 | 1·0 |
| Dury *et al.* (1998) | Water | 5–60[b] | 1000 | 0·07025 | 0·08–1·20 | 0·27 |
| | 2% Butanol | 5–60[b] | 996 | 0·04405 | 0·08–1·20 | 0·27 |
| | 6% Butanol | 5–60[b] | 989 | 0·03075 | 0·08–1·20 | 0·27 |
| Culligan *et al.* (2004) | Water–Soltrol | 72–105[c] | 1000 | 0·0378[d] | 0·6–1·4 | 0·85 |

[a] Effective diameter values used for comparisons against simulations
[b] Source: Morrow (1976)
[c] Source: Powers *et al.* (1996)
[d] Source: Schaap *et al.* (2007)

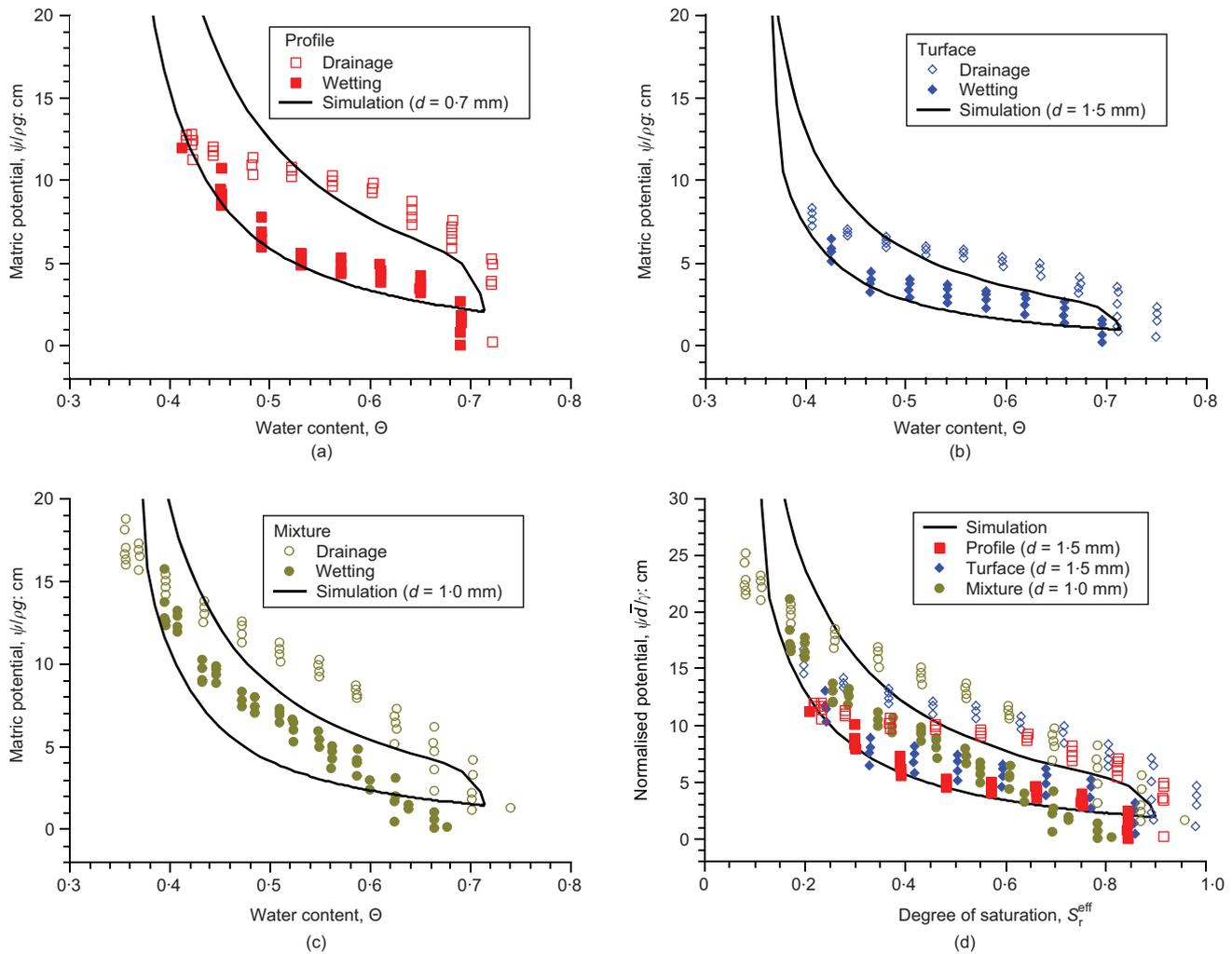

**Fig. 4.** Comparison of simulated water retention curves and experimental data: (a) profile; (b) turface; (c) mixture; (d) normalised data. The experimental data are normalised with the density $\rho$, effective grain size $\bar{d}$ and surface tension $\gamma$, listed in Table 2; values of $\Theta_{min} = 0.32$ and $\Theta_{max} = 0.75$ were used to convert the water content to the effective degree of saturation (Blonquist et al., 2006)

The predictions of water potential for the drainage branch at low saturation are usually higher than the measured values, possibly because the simulations were initially closely packed and thus inhibited grain motion when compared with the experimental systems.

In the second set of experiments, Dury et al. (1998) used various water–butanol mixtures at 0, 2 and 6% to alter the liquid surface tension. These wetting liquids were used to measure the corresponding capillary pressure–saturation relations. Similarly to Fig. 4, Figs 5(a)–5(c) show comparisons of predicted and experimental hysteresis in the space of matric potential and degree of saturation, again being normalised in Fig. 5(d). The calculated correlation coefficients were 0·93–0·98 for wetting branches and 0·82–0·98 for drainage.

In the third set of experiments used for comparison, Culligan et al. (2004) employed a water–Soltrol 220 system and glass beads. Unlike the previous two sets of experiments, where the void was occupied by air, here 'void' represents the volume taken by the oil phase in the oil–water system. As a result, the contact angle in this case is that between water and oil, and the corresponding value can be found in Table 2. Figure 6(b) shows a comparison of the model prediction with data from Culligan et al. (2004) for the oil–water system. Using x-ray micro-tomography, Culligan et al. (2004) further confirmed that, in the sample, the water saturation ($S_r$) and oil saturation ($1 - S_r$) levels were rather homogeneous along the gravity direction. From the modelling perspective, the main differences between the air–water and oil–water systems are

- advancing and receding contact angles
- surface tension between the wetting liquid and 'void'
- compressibility of the 'void', as in equation (4).

Using normalised capillary potential, Figs 6(a) and (b) demonstrate the difference by using corresponding advancing and receding angles for air–water and oil–water systems. The correlation coefficients are 0·82 for wetting and 0·93 for drainage.

*Grain-scale information*
The proposed method is advantageous in offering insights into local grain-scale information such as distributions of local water contents and potential during wetting–drainage cycles. Figure 7 presents six temporal snapshots of spatially distributed water content and the probability distribution of stress components, capillary and air stresses, for each cell for $0·1 \leq S_r \leq 0·9$. It is relevant to compare snapshot pairs (a) and (f), (b) and (e), and (c) and (d), corresponding to drainage and wetting, respectively. Overall, it is found that the probability of high capillary stresses is higher during drainage, which explains the difference in the overall matric potential. Furthermore, the entrapped particle-cell air pressure is only present at higher saturations, but is rather symmetrical during drainage and wetting, and can therefore not be attributed to explain the hysteresis. This is expected, since equation (4) is associated only with high degrees of saturation.

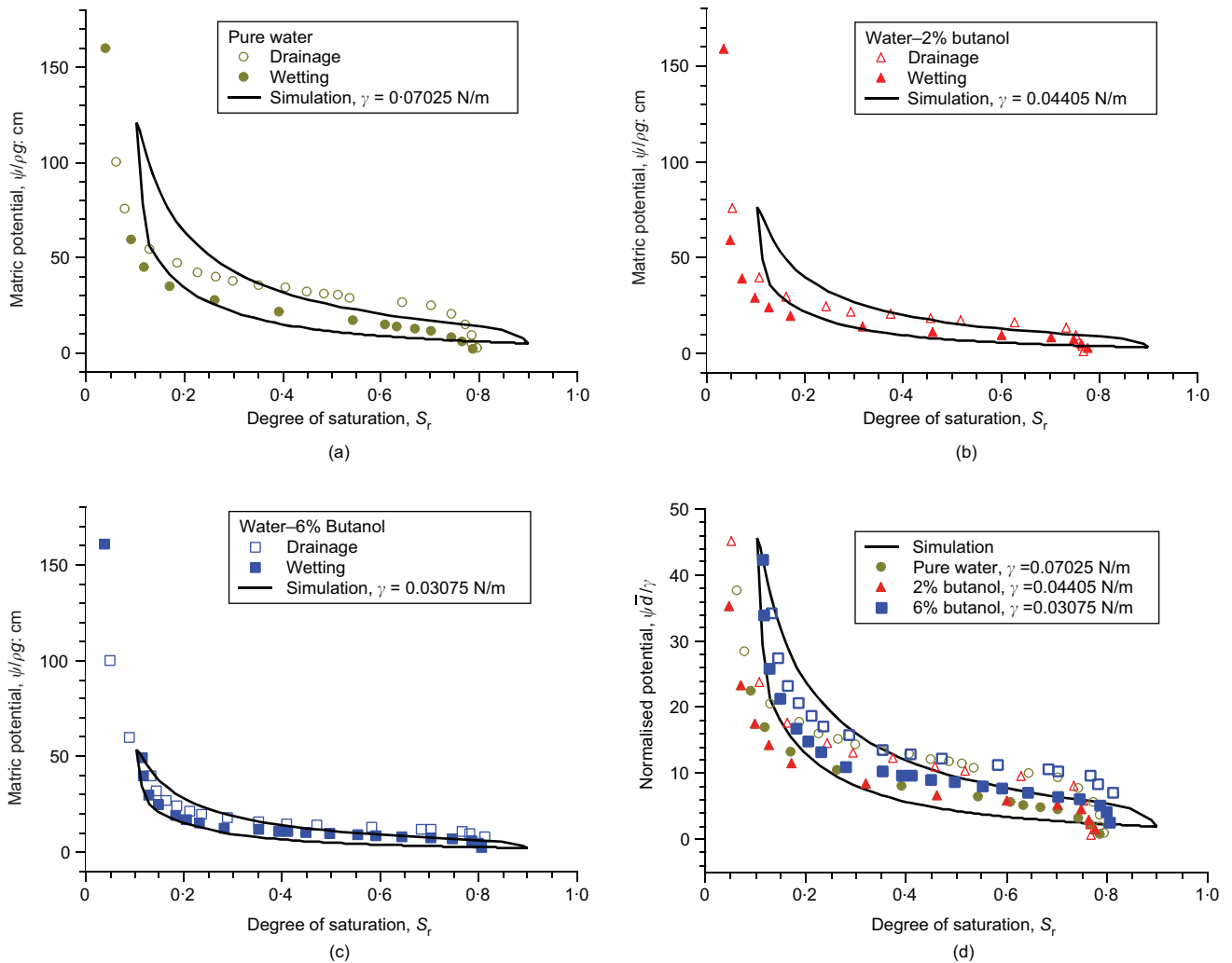

**Fig. 5.** Comparison of simulated retention curves and experimental data from Dury *et al.* (1998): (a) water; (b) 2% butanol; (c) 6% butanol; (d) normalised data. The experimental data are normalised with the density $\rho$, effective grain size $\bar{d}$ and surface tension $\gamma$, listed in Table 2

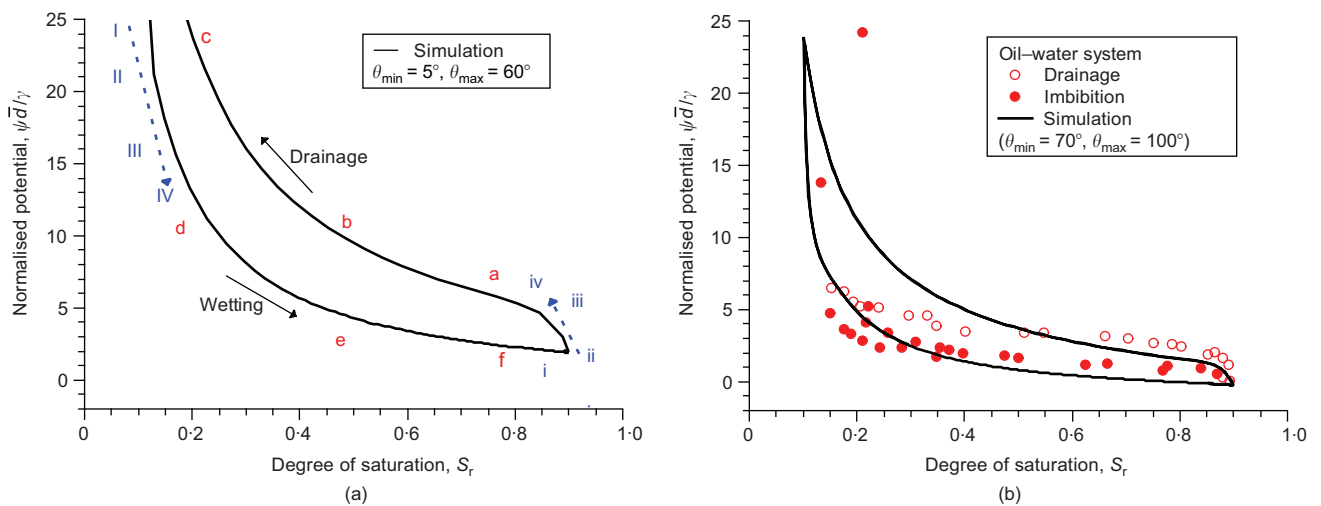

**Fig. 6.** Retention hysteresis curves with (a) $\theta_{min} = 5°$ and $\theta_{max} = 60°$ and (b) $\theta_{min} = 70°$ and $\theta_{max} = 100°$. In (a), the notation a–f indicates the cases for plotting water distribution in Fig. 7 and notation i to iv and I to IV for the cases in Fig. 8. In (b), the model prediction is compared with experimental data from Culligan *et al.* (2004) for an oil–water system

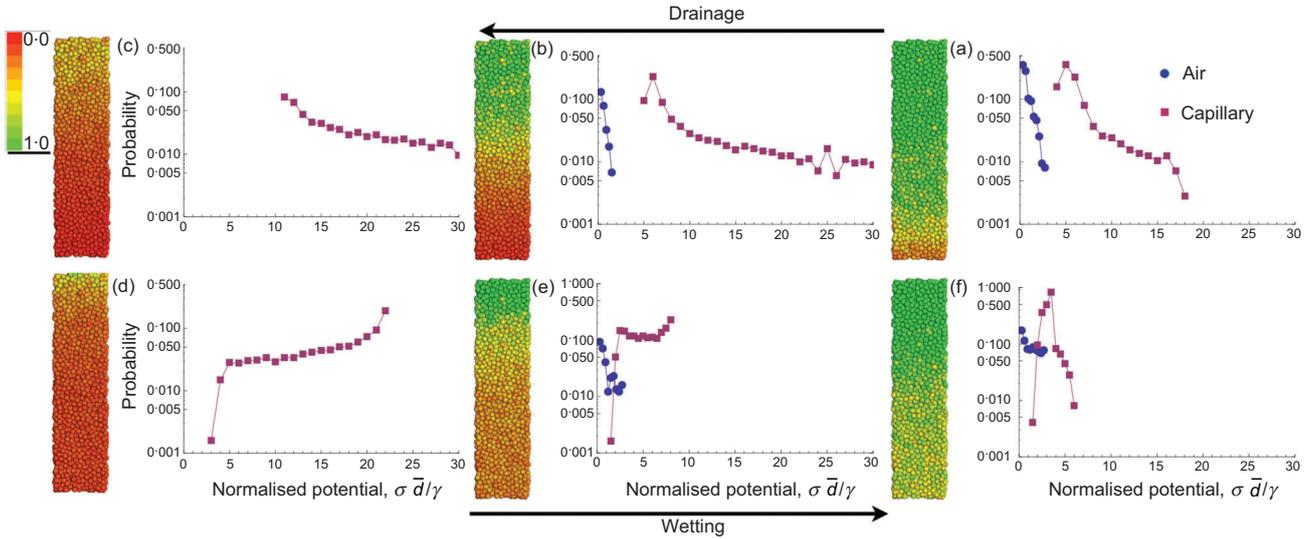

**Fig. 7.** Water distribution and grain-scale stress components in six stages – (a) to (c) for drainage and (d) to (f) for wetting – as indicated in Fig. 6(a). Each stage contains one profile and one probability distribution. Profiles show the local water distribution coloured by the degree of saturation with the scale between 0 and 1. Also shown are probability distributions of the magnitudes of particle-cell air and capillary stress components ($\sigma^{air}$ and $\sigma^{cap}$, respectively)

In this study, water retention hysteresis was connected to the contact angle dynamics via equation (3). Therefore, it is useful to plot the distributions of individual contact angles during wetting and drainage. Figure 8 shows the contact angle distribution and its evolution during transitions between drainage and wetting processes, at the stages shown in Fig. 6(a). This demonstrates how the distribution of contact angles inside the sample determines the overall water potential and thus provides an explanation for different water retention curves for wetting and drainage processes.

Future work will include sensitivity analyses on the effects of grain size distribution, porosity and hydraulic conductivity. The current study illustrates the role of contact angle dynamics in controlling the water retention hysteresis, but the method should enable study of the effects of grain motion and pore size distribution on retention hysteresis.

## CONCLUSIONS

A computational DEM model of the retention of fluids during wetting and drainage in partially saturated granular materials is proposed. By supporting this method with a new homogenisation scheme and employing grain-level quantities of water content and potential, this study demonstrates how small-scale hydro-mechanical parameters and variables control the phenomenology of larger scale water retention. Within the context of this study, a major finding was that the hysteresis of water retention during cycles of wetting and drainage was shown to arise from the dynamics of the solid–liquid contact angles as a function of the change in local water volumes. Comparison with experiments from different sources highlights the relative success of the model by directly involving only physically based micro-scale hydro-mechanical properties.

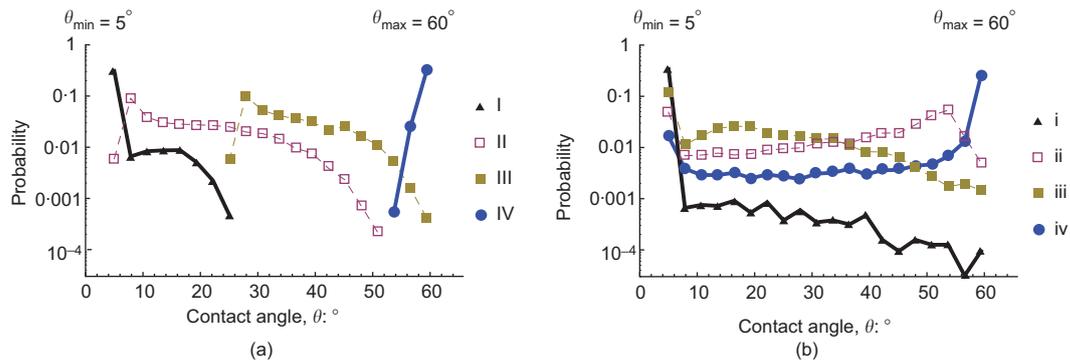

**Fig. 8.** Probability distribution of contact angle for individual capillary bridges during (a) transition from drainage to wetting and (b) from wetting to drainage. Stages I to IV and i to iv are identified in Fig. 6(a)

Acknowledgements
Financial support for this research from the Australian Research Council through grants DP120104926 and DE130101639 is greatly appreciated. The authors would like to thank Dr Jean-Michel Pereira for his critical comments. The DEM code and scripts for the present study can be downloaded from https://github.com/ganyx/CapDEM.

REFERENCES

Bachmann, J., Woche, S., Goebel, M., Kirkham, M. & Horton, R. (2003). Extended methodology for determining wetting properties of porous media. *Water Resources Res.* **39**, No. 12, paper 1353.

Blake, T. D. & Haynes, J. M. (1969). Kinetics of liquid/liquid displacement. *J. Colloid Interf. Sci.* **30**, No. 3, 421–423.

Blonquist, J. M., Jones, S. B., Lebron, I. & Robinson, D. (2006). Microstructural and phase configurational effects determining water content: Dielectric relationships of aggregated porous media. *Water Resources Res.* **42**, No. 5, 1–13.

Buscarnera, G. & Einav, I. (2012). The yielding of brittle unsaturated granular soils. *Géotechnique* **62**, No. 2, 147–160.

Chareyre, B., Cortis, A., Catalano, E. & Barthélémy, E. (2012). Pore-scale modeling of viscous flow and induced forces in dense sphere packings. *Transp. Porous Media* **94**, No. 2, 595–615.

Cho, G. & Santamarina, J. (2001). Unsaturated particulate materials-particle-level studies. *J. Geotech. Geoenviron. Engng* **127**, No. 1, 84–96.

Coussy, O., Pereira, J.-M. & Vaunat, J. (2010). Revisiting the thermodynamics of hardening plasticity for unsaturated soils. *Comput. Geotech.* **37**, No. 1–2, 207–215.

Crawford, J. W., Harris, J. A., Ritz, K. & Young, I. M. (2005). Towards an evolutionary ecology of life in soil. *Trends Ecol. Evol.* **20**, No. 2, 81–87.

Culligan, K. A., Wildenschild, D., Christensen, B. S. B., Gray, W. G., Rivers, M. L. & Tompson, A. F. B. (2004). Interfacial area measurements for unsaturated flow through a porous medium. *Water Resources Res.* **40**, No. 12, paper W12413.

Cundall, P. A. & Strack, O. D. L. (1979). A discrete numerical model for granular assemblies. *Géotechnique* **29**, No. 1, 47–65.

Di Renzo, A. & Di Maio, F. P. (2007). Homogeneous and bubbling fluidization regimes in DEMCFD simulations: hydrodynamic stability of gas and liquid fluidized beds. *Chem. Engng Sci.* **62**, No. 1–2, 116–130.

Dury, O., Fischer, U. & Schulin, R. (1998). Dependence of hydraulic and pneumatic characteristics of soils on a dissolved organic compound. *J. Contam. Hydrol.* **33**, No. 1–2, 39–57.

Gan, Y. & Kamlah, M. (2010). Discrete element modelling of pebble beds: with application to uniaxial compression tests of ceramic breeder pebble beds. *J. Mech. Phys. Solids* **58**, No. 2, 129–144.

Gerhard, J. I. & Kueper, B. H. (2003). Capillary pressure characteristics necessary for simulating DNAPL infiltration, redistribution, and immobilization in saturated porous media. *Water Resources Res.* **39**, No. 8, paper 1212.

Gili, J. A. & Alonso, E. E. (2002). Microstructural deformation mechanisms of unsaturated granular soils. *Int. J. Numer. Anal. Methods Geomech.* **26**, No. 5, 433–468.

Goebel, M. & Bachmann, J. (2004). Water potential and aggregate size effects on contact angle and surface energy. *Soil Sci. Soc. Am. J.* **68**, No. 2, 383–393.

Gras, J.-P., Delenne, J.-Y., Soulié, F. & El Youssoufi, M. (2011). DEM and experimental analysis of the water retention curve in polydisperse granular media. *Powder Technol.* **208**, No. 2, 296–300.

Gray, W. G. & Schrefler, B. A. (2001). Thermodynamic approach to effective stress in partially saturated porous media. *Euro. J. Mech. A Solids* **20**, No. 4, 521–538.

Gu, C., Maggi, F., Riley, W. J., Hornberger, G. M., Xu, T., Oldenburg, C. M., Spycher, N., Miller, N. L., Venterea, R. T. & Steefel, C. (2009). Aqueous and gaseous nitrogen losses induced by fertilizer application. *J. Geophys. Res.* **114**, No. G1, G01006.

Heinse, R., Jones, S. B., Steinberg, S. L., Tuller, M. & Or, D. (2007). Measurements and modeling of variable gravity effects on water distribution and flow in unsaturated porous media. *Vadose Zone J.* **6**, No. 4, 713–724.

Houlsby, G. T. (1997). The work input to an unsaturated granular material. *Géotechnique* **47**, No. 1, 193–196.

Jiang, M., Leroueil, S. & Konrad, J. (2004). Insight into shear strength functions of unsaturated granulates by DEM analyses. *Comput. Geotech.* **31**, No. 6, 473–489.

Laio, F., Porporato, A., Ridolfi, L. & Rodriguez-Iturbe, I. (2001). Plants in water-controlled ecosystems: active role in hydrologic processes and response to water stress: II. Probabilistic soil moisture dynamics. *Adv. Water Resources* **24**, No. 7, 707–723.

Lappalainen, K., Manninen, M., Alopaeus, V., Aittamaa, J. & Dodds, J. (2008). An analytical model for capillary pressure saturation relation for gas liquid system in a packed-bed of spherical particles. *Transp. Porous Media* **77**, No. 1, 17–40.

Liu, S. H. & Sun, D. A. (2002). Simulating the collapse of unsaturated soil by DEM. *Int. J. Numer. Anal. Methods Geomech.* **26**, No. 6, 633–646.

Loret, B. & Khalili, N. (2002). An effective stress elasticplastic model for unsaturated porous media. *Mech. Mater.* **34**, No. 2, 97–116.

Lourenço, S., Gallipoli, D., Augarde, C., Toll, D., Fisher, P. & Congreve, A. (2012). Formation and evolution of water menisci in unsaturated granular media. *Géotechnique* **62**, No. 3, 193–199.

Maggi, F. (2012). Multiphase capillary rise of multicomponent miscible liquids. *Coll. Surf. A: Physicochem. Engng Aspects* **415**, 119–124.

Maggi, F., Gu, C., Riley, W. J., Hornberger, G. M., Venterea, R. T., Xu, T., Spycer, N., Miller, N. L. & Oldenburg, C. M. (2008). A mechanistic treatment of the dominant soil nitrogen cycling processes: Model development, testing, and application. *J. Geophys. Res.: Biogeosci.* **113**, No. G2, G02016.

Magnuson, M. L. & Speth, T. F. (2005). Quantitative structure–property relationships for enhancing predictions of synthetic organic chemical removal from drinking water by granular activated carbon. *Environ. Sci. Technol.* **39**, No. 19, 7706–7711.

Morrow, N. (1976). Capillary pressure correlations for uniformly wetted porous media. *J. Can. Petrol. Technol.* **15**, No. 4, 49–69.

Nikooee, E., Habibagahi, G., Hassanizadeh, S. M. & Ghahramani, A. (2012). Effective stress in unsaturated soils: a thermodynamic approach based on the interfacial energy and hydromechanical coupling. *Transp. Porous Media* **96**, No. 2, 369–396.

Nuth, M. & Laloui, L. (2008). Advances in modelling hysteretic water retention curve in deformable soils. *Comput. Geotech.* **35**, No. 6, 835–844.

Pereira, J. & Arson, C. (2013). Retention and permeability properties of damaged porous rocks. *Comput. Geotech.* **48**, 272–282.

Pilon-Smits, E. (2005). Phytoremediation. *Ann. Rev. Plant Biol.* **56**, 15–39.

Powers, S., Anckner, W. & Seacord, T. (1996). Wettability of NAPL-contaminated sands. *J. Environ. Engng* **122**, No. 10, 889–896.

Revil, A. & Cathles, L. M. (1999). Permeability of shaly sands. *Water Resources Res.* **35**, No. 3, 651–662.

Richards, L. (1931). Capillary conduction of liquids through porous mediums. *Physics* **1**, No. 5, 318–333.

Russell, A. & Buzzi, O. (2012). A fractal basis for soil-water characteristics curves with hydraulic hysteresis. *Géotechnique* **62**, No. 3, 269–274.

Rycroft, C., Grest, G., Landry, J. & Bazant, M. (2006). Analysis of granular flow in a pebble-bed nuclear reactor. *Phys. Rev. E* **74**, No. 2, 021306.

Salt, D. E., Smith, R. D. & Raskin, I. (1998). Phytoremediation. *Ann. Rev. Plant Physiol. Plant Molec. Biol.* **49**, No. 1, 643–668.

Schaap, M. G., Porter, M. L., Christensen, B. S. B. & Wildenschild, D. (2007). Comparison of pressure-saturation characteristics derived from computed tomography and lattice Boltzmann simulations. *Water Resources Res.* **43**, No. 12, W12S06.

Schideman, L. C., Mariñas, B. J., Snoeyink, V. L. & Campos, C. (2006). Three-component competitive adsorption model for fixed-bed and moving-bed granular activated carbon adsorbers. Part I. Model development. *Environ. Sci. Technol.* **40**, No. 21, 6805–6811.

Scholtès, L., Chareyre, B., Nicot, F. & Darve, F. (2009a). Micromechanics of granular materials with capillary effects. *Int. J. Engng Sci.* **47**, No. 1, 64–75.

Scholtès, L., Hicher, P., Nicot, F., Chareyre, B. & Darve, F. (2009b). On the capillary stress tensor in wet granular materials. *Int. J. Numer. Anal. Methods Geomech.* **33**, No. 10, 1289–1313.

Schwarze, R., Gladkyy, A., Uhlig, F. & Luding, S. (2013). Rheology of weakly wetted granular materials – a comparison of experimental and numerical data. *Granular Matter*, in press.

Serrano, R., Iskander, M. & Tabe, K. (2011). 3D contaminant flow imaging in transparent granular porous media. *Géotechnique Lett.* **1**, No. 3, 71–78.

Sheng, D., Gens, A., Fredlund, D. & Sloan, S. (2008). Unsaturated soils: from constitutive modelling to numerical algorithms. *Comput. Geotech.* **35**, No. 6, 810–824.

Soulié, F., Cherblanc, F., El Youssoufi, M. & Saix, C. (2006a). Influence of liquid bridges on the mechanical behaviour of polydisperse granular materials. *Int. J. Numer. Anal. Methods Geomech.* **30**, No. 3, 213–228.

Soulié, F., El Youssoufi, M. S, Cherblanc, F. & Saix, C. (2006b). Capillary cohesion and mechanical strength of polydisperse granular materials. *Euro. Phys. J. E: Soft Matter* **21**, No. 4, 349–357.

Tindall, J. A., Kunkel, J. R. & Anderson, D. E. (1999). *Unsaturated zone hydrology for scientists and engineers*. Upper Saddle River, NJ: Prentice Hall.

van Genuchten, M. (1980). A closed-form equation for predicting the hydraulic conductivity of unsaturated soils. *Soil Sci. Soc. Am. J.* **44**, No. 5, 892–898.

Zeghal, M. & El Shamy, U. (2004). A continuum-discrete hydromechanical analysis of granular deposit liquefaction. *Int. J. Numer. Anal. Methods Geomech.* **28**, No. 14, 1361–1383.